\documentclass[conference]{IEEEtran}
\usepackage{cite}
\usepackage{epsfig}
\usepackage{epstopdf}
\usepackage{graphicx}
\usepackage{amsmath}
\usepackage{times}
\usepackage{amssymb}
\usepackage{framed}
\usepackage{graphicx,bm}
\usepackage{algorithm}
\usepackage{algorithmic}
\usepackage{amsfonts}
\usepackage{multicol}
\usepackage{psfrag}
\usepackage{subfigure}
\usepackage{footnote}
\usepackage{array}
\usepackage{setspace}
\usepackage{url}
\usepackage{float}
\usepackage{color}
\usepackage{cite}

\newtheorem{lemma}{\textbf{Lemma}}

\allowdisplaybreaks

\begin{document}
\title{One-Way URLLC with Truncated Channel Inversion Power Control}

\author{\IEEEauthorblockN{Chunhui Li\IEEEauthorrefmark{1}, Shihao Yan\IEEEauthorrefmark{2}, Nan Yang\IEEEauthorrefmark{1}, Xiangyun Zhou\IEEEauthorrefmark{1}, and Riqing Chen\IEEEauthorrefmark{3}}\\
\IEEEauthorblockA{\IEEEauthorrefmark{1}Research School of Electrical, Energy and Materials Engineering,\\ The Australian National University, Canberra, ACT 2601, Australia\\
\IEEEauthorrefmark{2}School of Engineering, Macquarie University, Sydney, NSW 2109, Australia\\
\IEEEauthorrefmark{3}School of Computer and Information Sciences, Fujian Agriculture and Forestry University, Fuzhou, China\\
Emails: \{chunhui.li, nan.yang, xiangyun.zhou\}@anu.edu.au, shihao.yan@mq.edu.au, riqing.chen@fafu.edu.cn}}


\maketitle

\begin{abstract}
In this work, we consider one-way ultra-reliable and low-latency communication (URLLC), where only the transmission in one direction requires URLLC and the transmission in the opposite direction does not. In order to meet the low-latency requirement of the one-way URLLC, we propose to use a truncated channel inversion power control (CIPC) to eliminate the requirement and the associated overhead of the training-based channel estimation at the receiver, while utilizing the multi-antenna technique at the transmitter to enhance the communication reliability. We first derive the transmission outage probability achieved by the truncated CIPC by considering the impact of a finite blocklength and a maximum transmit power constraint. Then, we determine the optimal constant power of the received signals in the truncated CIPC, which minimizes the transmission outage probability. Our examination shows that the proposed truncated CIPC is an effective means to achieve the one-way URLLC, where the tradeoff among reliability, latency, and required resources (e.g., the required number of transmit antennas, or the required maximum transmit power) is revealed.
\end{abstract}

\section{Introduction}
Ultra-reliable and low-latency communication (URLLC) is envisioned to support mission critical applications, e.g., industrial automation and remote surgery, where the requirements of latency and reliability are stringent. Specifically, in URLLC scenarios, the end-to-end delay and the decoding error probability are on the order of $1$ ms and $10^{-7}$, respectively~\cite{Li2018PROCEDIEEE}. Some fundamental aspects of URLLC have been studied in the literature (e.g., \cite{She2017MCOM,Popovski2018NETW,Ri2019WCOM,Hu2018NETW,Sun2018TWC}). Considering the low-latency constraint, the coding blocklength (i.e., channel uses or packet size) is required to be as short as possible in the context of URLLC applications~\cite{Johansson2015ICCW,Yilmaz2015ICCW}.

In practice, it is a big challenge to satisfy the quality-of-service (QoS) requirements (i.e., the ultra-reliable and low-latency requirements) when the coding blocklength becomes short and limited. Besides that the decoding error probability is no longer negligible for finite blocklength, another main reason is that it is hard to achieve accurate channel state information (CSI) in wireless networks within such a short time period. Existing works, aiming at ensuring the QoS requirements of URLLC in the finite blocklength regime, mainly assumed that the channel state information (CSI) is available or can be accurately estimated by using negligible channel uses. For example, radio resource management in the finite blocklength regime was examined to satisfy QoS requirement with signalling overhead, for downlink transmission via cross-layer resource allocation in \cite{She2018TWC}, and for short packet delivery via joint uplink and downlink optimization in \cite{She2018TCOM}. In \cite{Hu2018TWC}, the optimal power allocation was studied for QoS-constrained downlink multi-user networks in different types of data arrival. In these works, the cost of channel estimation in the context of satisfying QoS requirements was ignored by adopting the aforementioned assumption (i.e., CSI is perfectly available or estimated by using negligible resources). We note that the impact of channel estimation overhead on transmitting short packets in the finite blocklength regime was examined in \cite{Durisi2016TOC} and \cite{Li2019TVT}. However, as aforementioned, when the low-latency requirement is very stringent, we may not have any resource to conduct channel estimation.


The ultra-reliable requirement cannot be satisfied by retransmission or the transmission that requires the traditional channel estimation in URLLC scenarios. When channel reciprocity holds, channel inversion power control (CIPC) can be used for wireless communication, while eliminating the conventional requirement that a receiver should know CSI to conduct decoding \cite{ElSawy2014TCOM,ElSawy2014TWC}. This is due to that a transmitter can use CIPC to vary its transmit signal and power in order to ensure that the power of the received signals at the receiver is a constant value, which is \emph{a prior} agreed between the transmitter and receiver. We note that CIPC requires that CSI is available at a transmitter, but it avoids the cost of feeding CSI back from the transmitter to the receiver. This property leads to the fact that CIPC may serve as a key enabler of one-way URLLC in future wireless networks. Although CIPC has been studied in different communication scenarios (e.g.,  \cite{ElSawy2014TCOM,ElSawy2014TWC,Hu2018CIPC}), its performance and the associated optimization of the agreed constant power have never been investigated in the context of URLLC. This mainly motivates this work.

In this work, we adopt truncated CIPC to achieve one-way URLLC, where the maximum transmit power at the transmitter is considered. Specifically, we first derive the transmission outage probability achieved by the truncated CIPC by considering the impact of a finite blocklength and a maximum transmit power constraint. We then optimize the agreed constant power of the received signals to minimize this transmission outage probability. We note that one-way URLLC has a wide range of applications. For example, in vehicular wireless networks the communication from a vehicle to a base station that delivers warning information requires one-way URLLC, while the communication on the other way (mainly delivering videos or music for entertainment) may not require URLLC. Similar application scenarios can also be found in digital medical systems and industrial Internet of Things.

\section{System Model}
In this section, we first detail our considered scenario of one-way URLLC {together} with the adopted assumptions. Then, we explain our proposed scheme (i.e., the truncated CIPC) in details and present the definition of the resultant reliability outage probability.
\begin{figure}[!t]
    \begin{center}
        \includegraphics[width=0.8\columnwidth]{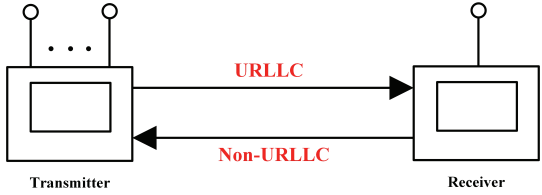}
        \caption{One-way URLLC.}\label{system_model}\vspace{-2em}
    \end{center}
\end{figure}

\subsection{Considered Scenario and Adopted Assumptions}

As shown in Fig.~\ref{system_model}, in this work we consider a one-way URLLC scenario in a time division duplex (TDD) multiple-input single-output (MISO) communications system, where an $N_{\mathrm{t}}$-antenna transmitter sends urgent information to a single-antenna receiver with the stringent requirement of latency and reliability. We denote $\mathbf{h}_{\mathrm{u}}$ as the $N_{\mathrm{t}} \times 1$ uplink channel vector from the receiver to the transmitter and denote $\mathbf{h}_{\mathrm{d}}$ as the $1 \times N_{\mathrm{t}}$ downlink channel vector from the transmitter to the receiver. As such, the downlink transmission considered in this work requires URLLC (the downlink transmission should occur within a finite blocklength $T$, i.e., $T$ channel uses), while the uplink transmission does not. All the channels are subject to independent quasi-static Rayleigh fading. We assume perfect channel reciprocity in this work, i.e., $\mathbf{h}_{\mathrm{u}}^{T}=\mathbf{h}_{\mathrm{d}}$ during one fading block, where $\mathbf{h}_{\mathrm{u}}^{T}$ denotes the transpose of $\mathbf{h}_{\mathrm{u}}$. The entries of each channel are assumed to be independent and identically distributed (i.i.d.) circularly symmetric complex Gaussian random variables with zero mean and unit variance, e.g., $\mathbf{h}_{\mathrm{d}}\sim\mathcal{CN} \left(0,\mathbf{I}_{N_{\mathrm{t}}}\right)$, where $\mathcal{CN}\left(\mu,\nu\right)$ denotes the complex Gaussian distribution with the mean of $\mu$ and the variance of $\nu$ and $\mathbf{I}_{N_{\mathrm{t}}}$ is an $N_{\mathrm{t}} \times N_{\mathrm{t}}$ identity matrix.

We further assume that the transmitter knows $\mathbf{h}_{\mathrm{u}}$ perfectly. This is due to the fact that the uplink transmission does not have strict requirement on delay, which makes it possible for the receiver to periodically broadcast pilots such that the transmitter can estimate $\mathbf{h}_{\mathrm{u}}$ perfectly. With the perfect channel reciprocity, the transmitter also knows $\mathbf{h}_{\mathrm{d}}$ perfectly. We note that the receiver does not know $\mathbf{h}_{\mathrm{u}}$ or $\mathbf{h}_{\mathrm{d}}$, since there is no feedback from the transmitter to the receiver. In the downlink communication, all the channel uses are for data transmission when urgent information is on demand to transmit. This will significantly reduce the communication latency and improve transmission reliability, in order to meet the requirements of URLLC. To enable the receiver to decode the information without knowing the accurate CSI, the truncated CIPC (i.e., channel inversion power control) will be used at the transmitter based on the perfectly known $\mathbf{h}_{\mathrm{d}}$, which will be detailed in the following subsection.

\subsection{Truncated Channel Inversion Power Control}

In this work, we consider the truncated CIPC at the transmitter to enable the receiver to decode received signals without knowing $\mathbf{h}_{\mathrm{d}}$. The received signal in one channel use is given by
\begin{align}\label{recei_signal}
y=\sqrt{P_{\mathrm{a}}}\mathbf{h}_{\mathrm{d}}\mathbf{x}+w,
\end{align}
where $w$ is the AWGN at the receiver with zero mean and variance $\sigma_{w}^{2}$, $\mathbf{x}$ is the transmitted signal, which is subject to the average power constraint, i.e.,  $\mathbb{E}\left[\|\mathbf{x}\|^{2}\right]=1$ with $\mathbb{E}\left[\cdot\right]$ denoting expectation, and $P_{\mathrm{a}}$ is the transmit power.
In order to counteract the impact of the phase in the downlink channel at the receiver, the transmitted signal $\mathbf{x}$ is designed as
\begin{align}\label{trans_signal}
\mathbf{x}=\frac{\mathbf{h}^{\dag}_{\mathrm{d}}}{\|\mathbf{h}_{\mathrm{d}}\|}u,
\end{align}
 where $u$ is the information signal transmitted from the transmitter to the receiver. Following \eqref{recei_signal} and \eqref{trans_signal}, the signal-to-noise ratio (SNR) at the receiver can be written as $\gamma={P_{\mathrm{a}} \|\mathbf{h}_{\mathrm{d}}\|^{2}}/{\sigma_{w}^{2}}$.
In order to counteract the impact of the downlink channel gain at the receiver, in CIPC the transmitter varies its transmit power as per $\|\mathbf{h}_{\mathrm{d}}\|$, such that
\begin{align}\label{constant_Q}
P_{\mathrm{a}}\|\mathbf{h}_{\mathrm{d}}\|^{2}=Q,
\end{align}
where $Q$ is a pre-determined constant value \textit{a priori} agreed between the transmitter and receiver. Then, the SNR at the receiver can be rewritten as
\begin{align}
\gamma=\frac{Q}{\sigma_{w}^{2}}.
\end{align}

Considering Rayleigh fading for $\mathbf{h}_{\mathrm{d}}$, as per \eqref{constant_Q} we can see that the transmit power $P_{\mathrm{a}}$ may be infinite to guarantee $P_{\mathrm{a}}\|\mathbf{h}_{\mathrm{d}}\|^{2}=Q$ for some realizations of $\mathbf{h}_{\mathrm{d}}$, which is not practical. As such, in this work
we consider the truncated CIPC, where the transmitter is subject to a maximum transmit power constraint~\cite{ElSawy2014TWC}. Specifically, the transmitter only transmits information to the receiver when the downlink channel gain (i.e., $\|\mathbf{h}_{\mathrm{d}}\|^{2}$) is greater than some specific value. Mathematically, the transmit power is given by
\begin{align}\label{eq:TransmitPowerAP}
P_{\mathrm{a}} &=
     \begin{cases}
        \frac{Q}{\|\mathbf{h}_{\mathrm{d}}\|^{2}} ,& \|\mathbf{h}_{\mathrm{d}}\|^{2} \geq \frac{Q}{P_{\mathrm{max}}}  \\
        0 ,& \|\mathbf{h}_{\mathrm{d}}\|^{2} < \frac{Q}{P_{\mathrm{max}}},
       \end{cases}
\end{align}
where $P_{\mathrm{max}}$ is the maximal transmit power. Based on \eqref{eq:TransmitPowerAP}, we can see that the transmitter does not always transmit information to the receiver due to the maximum transmit power constraint. As such, in addition to the finite blocklength, the maximum transmit power is another factor that causes transmission outage, which should be minimized. Therefore, in the following subsection we present the definition of the transmission outage probability, which is used as the performance metric for our proposed truncated CIPC.

\subsection{Performance Metric: Transmission Outage Probability}

In this subsection, we define the transmission outage probability that is used to evaluate the performance of our proposed truncated CIPC scheme.

For a finite blocklength the decoding error probability at the receiver is not negligible, of which an asymptotic expression is given by \cite{Polyanskiy2010}
\begin{align}\label{eq:epsilon}
\epsilon (Q) &= f\left(\frac{\log_{2}(1+\gamma)- R}{\sqrt{V/T}}\right),
\end{align}
where $R$ is the information transmission rate, $V=(\log_{2}e)^2 \left[1-1/(1+\gamma)^2\right]$ is the channel dispersion, $f(\cdot)$ denotes the Q-function where $f(x)=\int_{x}^{\infty}e^{-t^2/2}/\sqrt{2\pi}dt$. We note that the decoding error probability given in \eqref{eq:epsilon} is non-zero due to the non-zero property of the Q-function.

As per \eqref{eq:TransmitPowerAP}, the probability that the transmit power is not zero, i.e., the probability that the transmitter sends information to the receiver, is given by
\begin{align}\label{trans_prob}
p_t(Q)=\mathrm{Pr}\left\{P_{\mathrm{a}} \leq P_{\mathrm{max}} \right\}.
\end{align}
We note that the transmission outage is not only caused by the maximum transmit power constraint. When the transmitter can guarantee $P_{\mathrm{a}}\|\mathbf{h}_{\mathrm{d}}\|^{2}=Q$, an outage can still occur due to the non-zero decoding errors in the finite blocklength regime. Therefore, the overall transmission outage probability for our considered truncated CIPC is given by
\begin{align}\label{outage_definition}
P_{\epsilon}(Q) = \epsilon (Q) p_t(Q) + (1-p_t(Q)).
\end{align}
We note that, although $\epsilon (Q)$ in \eqref{outage_definition} is conditioned on that the transmit power is not zero, it is still for a fixed SNR determined by $Q$, since in the CIPC the SNR is a constant, which does not vary with the channel gain. We also note that, for fixed $R$ and $T$, this transmission outage probability $P_{\epsilon}(Q)$ given in \eqref{outage_definition} depends on $Q$ heavily. Intuitively, there exists an optimal value of $Q$ that minimizes $P_{\epsilon}(Q)$, since $p_t(Q)$ monotonically decreases with $Q$ and $\epsilon (Q)$ decreases with $Q$. Therefore, in the following section we first derive a closed-form expression for $P_{\epsilon}(Q)$ and then we determine this optimal value of $Q$ in order to minimize $P_{\epsilon}(Q)$.

\section{Performance Analysis and Optimization Framework for the Truncated CIPC Scheme}

In this section, we analyze the transmission outage probability of the truncated CIPC scheme, based on which we develop a framework to optimize the value of $Q$ in order to improve its performance in the context of URLLC.

\subsection{Transmission Outage Probability Expression}
In the following lemma, we derive a closed-form expression for the transmission outage probability of the truncated CIPC scheme.
\begin{lemma}\label{lemma1}
The transmission outage probability of the truncated CIPC scheme in the context of URLLC is derived as
\begin{align}\label{eq:CloseFormPrEpsilon}
P_{\epsilon}(Q)
    &= 1-\left[1-\frac{\gamma\left(N_{\mathrm{t}},\frac{Q}{P_{\mathrm{max}}}\right)}{\Gamma(N_{\mathrm{t}})}\right]
    \notag\\
    &\times   \left[1-f\left(\frac{\sqrt{T}\left(\ln(1+Q/\sigma_{w}^{2})-R\ln2 \right)}{\sqrt{1-\frac{1}{(1+Q/\sigma_{w}^{2})^2}}}\right)\right],
\end{align}
where $\gamma(s,x)=\int_{0}^{x} t^{s-1}e^{-t} dt$ is the lower incomplete gamma function and $\Gamma(x)=(x-1)!$ is the gamma function.
\end{lemma}
\begin{IEEEproof}
In order to prove Lemma~\ref{lemma1}, we have to derive the expression of $P_{\epsilon}(Q)$ given in \eqref{outage_definition} by deriving the explicit expressions for $p_t(Q)$ and $\epsilon(Q)$. We first tackle the probability $p_t(Q)= \mathrm{Pr}\left\{P_{\mathrm{a}} \leq P_{\mathrm{max}} \right\}$.
Substituting \eqref{eq:TransmitPowerAP} into \eqref{trans_prob}, we have
\begin{align}\label{eq:ProofP0}
p_t(Q)
 &= 1\!-\!\mathrm{Pr}\left\{\!\|\mathbf{h}_{\mathrm{d}}\|^{2} \!\leq\! \frac{Q}{P_{\mathrm{max}}}\! \right\}
 =1\!-\!\frac{\gamma\left(\!N_{\mathrm{t}},\frac{Q}{P_{\mathrm{max}}}\!\right)}{\Gamma(N_{\mathrm{t}})},
\end{align}
where $f_{X}(x)=x^{N_\mathrm{t}-1}e^{-x}/{\Gamma(N_\mathrm{t})}$ and $F_{X}(x)=\gamma\left(N_{\mathrm{t}},x\right)/\Gamma(N_{\mathrm{t}})$ are the probability density function (pdf) and cumulative distribution function (cdf) of $\|\mathbf{h}_{\mathrm{d}}\|^{2}$, respectively.

Substituting $\gamma=Q/\sigma_{w}^{2}$ into \eqref{eq:epsilon}, the decoding error probability can be rewritten as
\begin{align}\label{eq:ProofEpsilon}
\epsilon(Q)
 &=f\left(\frac{\sqrt{T}\left[\ln(1+Q/\sigma_{w}^{2})-R\ln2 \right]}{\sqrt{1-\frac{1}{(1+Q/\sigma_{w}^{2})^2}}}\right).
\end{align}
Finally, substituting \eqref{eq:ProofP0} and \eqref{eq:ProofEpsilon}
into \eqref{outage_definition}, we obtain the desired result in \eqref{eq:CloseFormPrEpsilon}, which completes the proof.
\end{IEEEproof}

We note that the transmission outage probability $P_{\epsilon}(Q)$ is a monotonically increasing function of the transmission rate $R$, since $\epsilon(Q)$ monotonically increases with $R$ while $p_t(Q)$ is not a function of $R$. Meanwhile, $P_{\epsilon}(Q)$ monotonically decreases with $P_{\mathrm{max}}$, as  $p_t(Q)$ increases with $P_{\mathrm{max}}$, while $\epsilon(Q) <1$ does not depend on $P_{\mathrm{max}}$. In Section \ref{sec:numerical results}, we will examine what is the required maximum transmit power in order to achieve URLLC with a certain transmission rate and a maximum allowable transmission outage probability.

\subsection{Optimization Framework of $Q$}\label{sec:problem_formulation}

In this subsection, we focus on determining the optimal value of $Q$ to minimize the transmission outage probability $P_{\epsilon}(Q)$ for given $T$, $R$ and $P_{\mathrm{max}}$. Then, the optimization problem at the transmitter is given by
\begin{subequations}\label{Objective Function}
\begin{align}
\min\limits_{Q}&~P_{\epsilon}(Q)\\
\mathrm{s.t.}&~~R \leq \log_2(1 + Q/\sigma_w^2),\label{contraints:R}
\end{align}
\end{subequations}
where \eqref{contraints:R}  is the transmission rate constraint (i.e., the transmission rate should be no larger than the corresponding Shannon capacity).

\begin{figure}[!t]
    \begin{center}
        \includegraphics[height=2.7in]{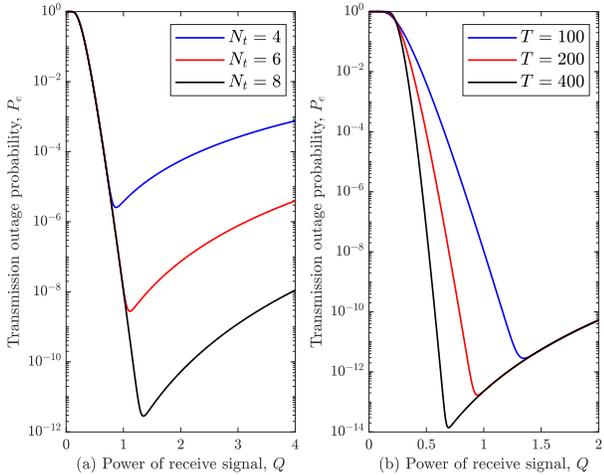}
        \caption{The transmission outage probability $P_{\epsilon}(Q)$ versus the constant value $Q$ in the truncated CIPC scheme for different values of $N_{\mathrm{t}}$ and $T$ with $R=0.3$, $P_{\textrm{max}}=10~dB$.}\label{Fig:1}\vspace{-2em}
    \end{center}
\end{figure}

Due to the high complexity of the expression for $P_{\epsilon}(Q)$ derived in Lemma~\ref{lemma1}, it is hard to analytically solve the optimization problem in \eqref{Objective Function}. We present the following lemma to aid numerically solving this optimization problem.

\begin{lemma}\label{lemma2}
{The transmission outage probability $P_{\epsilon}(Q)$ of the truncated CIPC is a convex function of $Q$ when $Q_0  < Q < P_{\mathrm{max}}(N_{\mathrm{A}}-1)$, where $Q_0$ is the solution of $\frac{\ln(1+Q)}{(1+Q)^{2}-1}=\frac{1}{3}$.}
\end{lemma}
\begin{IEEEproof}
See Appendix A.
\end{IEEEproof}



\section{Numerical Results}\label{sec:numerical results}

In this section, we present numerical results to examine the performance of the proposed truncated CIPC scheme in the context of URLLC, based on which we draw useful insights on the impact of some system parameters on the considered one-way URLLC.

In Fig.~\ref{Fig:1}, we plot the transmission outage probability $P_{\epsilon}(Q)$ of the truncated CIPC scheme versus different values of $Q$. In this figure, we first observe that there indeed exist an optimal value of $Q$ that minimizes $P_{\epsilon}(Q)$. We also observe that this optimal value is within the interval $(Q_0, P_{\mathrm{max}}(N_{\mathrm{t}}-1))$, which demonstrates the correctness of our Lemma~\ref{lemma2}.
In addition, we observe that the minimum value of $P_{\epsilon}(Q)$ significantly depends on the values of $N_{\mathrm{t}}$ and $T$, i.e., this minimum value decreases with $N_{\mathrm{t}}$ or $T$.  This first indicates that the reliability in URLLC can be improved by using more antennas in the truncated CIPC scheme. We note that without the considered CIPC scheme, increasing transmit antenna number may not improve reliability in URLLC, since the traditional channel estimation cost also increases as the number of transmit antennas increases. In Fig.~\ref{Fig:1}(a), we observe that, in the low regime of $Q$, $P_{\epsilon}(Q)$ for different values of $N_{\mathrm{t}}$ is almost the same. This is due to the fact that under this case $P_{\epsilon}(Q)$ is dominated by the decoding error probability $\epsilon(Q)$, which is not a function of $N_{\mathrm{t}}$. Meanwhile, in the high regime of $Q$, $P_{\epsilon}(Q)$ is different for different values of $\epsilon(Q)$, where $P_{\epsilon}(Q)$ is dominated by the probability that the transmitter sends information, which highly depends on $N_{\mathrm{t}}$. Similar observations and explanations can be applied to Fig.~\ref{Fig:1}(b).


\begin{figure}[!t]
    \begin{center}
        \includegraphics[height=2.7in]{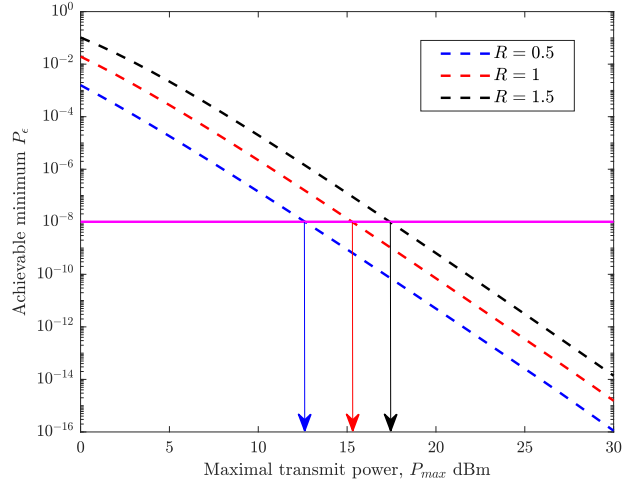}
        \caption{The minimum transmission outage probability $P_{\epsilon}^\ast(Q)$ versus the maximum transmit power $P_{\textrm{max}}$ for different values of $R$, where $N_{\mathrm{t}}=5$, $T=150$.}\label{Fig:2}\vspace{-2em}
    \end{center}
\end{figure}

In Fig.~\ref{Fig:2}, we plot the minimum transmission outage probability, denoted by $P_{\epsilon}^\ast(Q)$, achieved by the optimal $Q$ in the truncated CIPC, versus the maximum transmit power $P_{\textrm{max}}$ for different values of $R$. As expected, in this figure we first observe that $P_{\epsilon}^\ast(Q)$ monotonically decreases with $P_{\textrm{max}}$, since increasing $P_{\textrm{max}}$ can definitely increase $p_t(Q)$ for a fixed $\epsilon(Q)$. This demonstrates that the maximum transmit power plays a critical role in the truncated CIPC, as it determines the specific channel gain when the transmitter can conduct URLLC to the receiver. This figure demonstrates that, to guarantee a certain reliability, the required value of $P_{\textrm{max}}$ increases with the transmission rate $R$. In addition, we observe that $P_{\epsilon}^\ast(Q)$ increases as $R$ increases, which demonstrates the tradeoff between the transmission rate $R$ and the reliability.

\begin{figure}[!t]
    \begin{center}
        \includegraphics[height=2.7in]{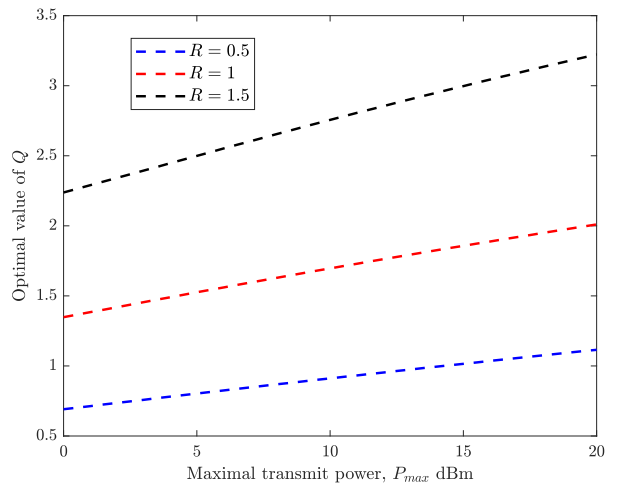}
        \caption{The optimal power of receive signal $Q$ versus the maximum transmit power $P_{\textrm{max}}$ for different values of $R$ with $N_{\mathrm{t}}=4$ and $T=200$.}\label{Fig:3}\vspace{-2em}
    \end{center}
\end{figure}

In Fig.~\ref{Fig:3}, we plot the optimal value of $Q$, which minimizes the transmission outage probability $P_{\epsilon}(Q)$, versus the maximum transmit power $P_{\textrm{max}}$ for different values of the transmission rate $R$. In this figure, we first observe the optimal $Q$ increases with $P_{\textrm{max}}$. Similarly, in this figure we also observe that the optimal $Q$ increases as the transmission rate $R$ increases. This is due to the fact that, as $R$ increases, the decoding error probability $\epsilon(Q)$ significantly increases, which again becomes the key factor limiting the overall transmission reliability, where we have to increase $Q$ to reduce this $\epsilon(Q)$.



\section{Conclusion}
In this work, we proposed using the truncated CIPC to achieve one-way URLLC in the MISO system. We proved that the achievable transmission outage probability is a convex function of the constant received signal power, i.e., $Q$ within a specific range. Based on that, we investigated the optimal value of $Q$ to minimize the achievable transmission outage probability by considering a finite blocklength and a maximum transmit power constraint. Our outcomes provide useful guidelines to assist the URLLC designers with the fundamental problem of determining the minimum required transmit power to guarantee a target reliability with a certain transmission rate.


\section*{Acknowledgements}
This work was supported by the ARC Discovery Project (DP180104062) and in part by the National Natural Science Foundation of China (Grant No. 61972093).

\appendices
\section{Proof of Lemma~\ref{lemma2}}\label{Proof:PrEpsilon_vs_Q}
In order to prove Lemma~\ref{lemma2}, we analyze the monotonicity and concavity of $P_{\epsilon}(Q)$ with respect to (w.r.t.) $Q$. We first derive the first-order derivative of $P_{\epsilon}(Q)$ w.r.t. $Q$ as
\begin{align}\label{fod:P_epsilon}
\frac{\partial P_{\epsilon}(Q) }{\partial Q}
  &=\frac{\partial \{p_t(Q)\}}{\partial Q}\biggl(\epsilon(Q)-1\biggr)+ p_t(Q) \frac{\partial \{\epsilon(Q)\}}{\partial Q}.
\end{align}
Then, the second-order derivative of $P_{\epsilon}(Q)$ w.r.t. $Q$ can be obtained as
\begin{align}\label{sod:P_epsilon}
\frac{\partial^2 P_{\epsilon}(Q) }{\partial Q^2}
  &=\frac{\partial^2 \{p_t(Q)\}}{\partial Q^2}\biggl(\epsilon(Q)-1\biggr)
  + 2 \frac{\partial \{p_t(Q)\}}{\partial Q} \frac{\partial \{\epsilon(Q)\}}{\partial Q}
  \notag\\
  &~~~~~~~~~+p_t(Q) \frac{\partial^2 \{\epsilon(Q)\}}{\partial Q^2}.
\end{align}

\begin{figure*}
\begin{align}\label{sod:Epsilon}
\frac{\partial^2 \epsilon(Q)}{\partial Q^2}
 &=\frac{A(Q)}{\sqrt{2\pi}}
 \exp\left(-\frac{A^{2}(Q)}{2}\right) \left\{\frac{\partial  \{A(Q)\} }{\partial Q}\right\}^{2}
 -\frac{1}{\sqrt{2\pi}} \exp\left(-\frac{A^{2}(Q)}{2}\right) \frac{\partial^2  \{A(Q)\} }{\partial Q^2},
\end{align}\vspace{-0.6cm}
\end{figure*}
\begin{figure*}
\begin{align}\label{sod:A(Q)}
\frac{\partial^2  \{A(Q)\} }{\partial Q^2}
    &=\frac{\sqrt{T}}{(1+Q)}
    \Biggl[
    \underbrace{\frac{3\left[\ln(1+Q)-R \ln2 \right]}{\left(\sqrt{(1+Q)^{2}-1}\right)^{5}}}_{\Psi_{1}}
    +\underbrace{\frac{3\left[\ln(1+Q)-R \ln2 \right]}{ \left(\sqrt{(1+Q)^{2}-1}\right)^{3}}}_{\Psi_{2}}
    -\underbrace{\frac{2}{\left(\sqrt{(1+Q)^{2}-1}\right)^{3}}}_{\Psi_{3}}
    -\underbrace{\frac{1}{\sqrt{(1+Q)^{2}-1}}}_{\Psi_{4}}
    \Biggr].
\end{align}\hrulefill
\vspace{-0.6cm}
\end{figure*}

To find the sign of $\frac{\partial^2 P_{\epsilon}(Q)}{\partial Q^2}$, we first need to address $\frac{\partial p_t(Q)}{\partial Q}$ and $\frac{\partial^2 p_t(Q)}{\partial Q^2}$. According to \eqref{eq:ProofP0}, the first-order derivative of $p_t(Q)$ w.r.t. $Q$ is given by
\begin{align}\label{fod:P_t}
\frac{\partial p_t(Q)}{\partial Q}
 &=-\frac{e^{-\frac{Q}{P_{\mathrm{max}}}} (\frac{Q}{P_{\mathrm{max}}})^{N_{\mathrm{A}}-1}}{P_{\mathrm{max}}\Gamma(N_{\mathrm{A}})}<0.
\end{align}
We find that $p_t(Q)$ is a monotonically decreasing function of $Q$ due to $\frac{\partial p_t(Q)}{\partial Q}<0$. The second-order derivative of $p_t(Q)$ w.r.t. $Q$ is given by
\begin{align}\label{sod:P_t}
\frac{\partial^2 p_t(Q)}{\partial Q^2}
 &=\frac{e^{-\frac{Q}{P_{\mathrm{max}}}} \!(\frac{Q}{P_{\mathrm{max}}}\!)^{N_{\mathrm{A}}+1} \biggl[\!Q-P_{\mathrm{max}}(N_{\mathrm{A}}-1)\!\biggr]}{Q^3 \Gamma(N_{\mathrm{A}})}.
\end{align}
The sign of $\frac{\partial^2 p_t(Q)}{\partial Q^2}$ has three outcomes, which are
\begin{align}
\frac{\partial^2 p_t(Q)}{\partial Q^2} &
     {
     \begin{cases}
        < 0 ,& 0 < Q < P_{\mathrm{max}}(N_{\mathrm{A}}-1) \\
        =0 ,& Q=P_{\mathrm{max}}(N_{\mathrm{A}}-1)\\
        > 0 ,& Q > P_{\mathrm{max}}(N_{\mathrm{A}}-1).
       \end{cases}
       }
\end{align}

{It is worth mention that $Q$ should be in the range of $(0, P_{\mathrm{max}}(N_{\mathrm{A}}-1))$ for arbitrary $N_{\mathrm{t}}$.} It is due to the fact that
when $Q = P_{\mathrm{max}}(N_{\mathrm{t}}-1)$, as per \eqref{eq:ProofP0} the probability $p_t(Q)$ becomes a function of only the variable $N_{\mathrm{t}}$, which is
\begin{align}
\dot{p_t}(N_{\mathrm{t}})  = \dot{p_t}(P_{\mathrm{max}}(N_{\mathrm{t}}-1)) =1- \frac{\gamma\left(N_{\mathrm{t}},{N_{\mathrm{t}}-1}\right)}{\Gamma\left(N_{\mathrm{t}}\right)}.
\end{align}
We note that $\dot{p_t}(N_{\mathrm{t}})$ is a monotonically {decreasing} function of $N_{\mathrm{t}}$ and thus $1 - \dot{p_t}(N_{\mathrm{t}})$ {increases and tends to a constant value (i.e., $0.5$) with $N_{\mathrm{t}}$.
However, we note that using more transmit antennas will not be beneficial to improve reliability when $Q=P_{\mathrm{max}}(N_{\mathrm{A}}-1)$. Following \eqref{outage_definition}, we have $P_{\epsilon}(Q) > 1 - p_t(Q)$. As such, for given $P_{\mathrm{max}}$, we cannot meet the ultra-reliable requirement of URLLC by setting $Q = P_{\mathrm{max}}(N_{\mathrm{t}}-1)$ and we have to decrease $Q$ in order to further increase the value of $p_t(Q)$. Therefore, reducing the value of $Q$ is the only solution to guarantee $1 - \dot{p_t}(N_{\mathrm{t}}) \leq 10^{-7}$.
We also note that $p_t(Q)=1-{\gamma\left(\!N_{\mathrm{t}},\frac{Q}{P_{\mathrm{max}}}\!\right)}/
{\Gamma(N_{\mathrm{t}})}$ is a monotonically decreasing function of $Q$ due to $\frac{\partial p_t(Q)}{\partial Q}<0$ proved in \eqref{fod:P_t}. Thus, for $Q > P_{\mathrm{max}}(N_{\mathrm{A}}-1)$, the term $1-p_t(Q)$ will be larger than $0.5$ which also violates the requirement of URLLC.}

Then, we calculate the first-order partial derivative of $\epsilon(Q)$ w.r.t. $Q$, which is given by
\begin{align}\label{fod:Epsilon}
\frac{\partial \{\epsilon(Q)\}}{\partial Q}
 &=-\frac{1}{\sqrt{2\pi}}\exp\left(-\frac{A^{2}(Q)}{2}\right) \frac{\partial  \{A(Q)\} }{\partial Q},
\end{align}
where we set {$\sigma_w^2=1$ for simplifying the analysis as the value of $\sigma_w^2$ does not affect the result}, which makes $\gamma=Q$ and $A(Q)={\sqrt{T}\left[\ln(1+Q)-R\ln2 \right]}/{\sqrt{1-{1}/{(1+Q)^2}}}$.

The second-order partial derivative of $\epsilon(Q)$ w.r.t. $Q$ (i.e., $\frac{\partial^2 \epsilon(Q)}{\partial Q^2}$) is given in \eqref{sod:Epsilon} , where the first-order partial derivative of $A(Q)$ w.r.t. $Q$ is given by
\begin{align}\label{fod:A(Q)}
\frac{\partial \{A(Q)\}}{\partial Q}
 &= \frac{\sqrt{T} \left[1-\frac{\ln(1+Q)-R \ln2}{(1+Q)^{2}-1}\right]}{\sqrt{(1+Q)^{2}-1}}.
\end{align}

Determining the sign of $\frac{\partial \{\epsilon(Q)\}}{\partial Q}$ is equivalent to figure out the sign of $\frac{\partial \{A(Q)\}}{\partial Q}$. To address this issue, we first define a function as
\begin{align}\label{eq:G(x)}
G(x)&=\frac{\ln {x}}{x^{2}-1},
\end{align}
where $x=1+Q$, and $x>1$ due to $Q>0$.

We note that ${d \{G(x)\}}/{d x}$ is given by
\begin{align}\label{fod:G(x)}
\frac{d \{G(x)\}}{d x}
 &= \frac{x-2x\ln{x}-\frac{1}{x}}{(x^{2}-1)^{2}}
 = \frac{g(x)}{(x^{2}-1)^{2}},
\end{align}
where $g(x)=x-2x\ln{x}-{1}/{x}$.

We find that the sign of $\frac{d \{G(x)\}}{d x}$ depends on $g(x)$ when $x>1$. It is clear that the first-order derivative of $g(x)$ w.r.t. $x$ is negative, where $\frac{d \{g(x)\}}{d x}=-\left(1-\frac{1}{x^{2}}\right)-2\ln{x}<0$ for $x>1$. In other words, $g(x)$ decreases with $x$ when $x>1$. As such, we can obtain that $g(x)<g(1)=0$. Thus, we can obtain that $\frac{d \{G(x)\}}{d x}<0$ for $x>1$, which means that $G(x)$ is a decreasing function w.r.t. $x$ for $x>1$.
As per L'Hospital's rule, we derive
\begin{align}
\lim_{x\to 1}G(x)=\lim_{x\to 1}\frac{d \{\ln {x}\}/dx}{d\{x^{2}-1\}/dx}=\lim_{x\to 1}\frac{1}{2x^{2}}=\frac{1}{2},
\\
\lim_{x\to\infty}G(x)=\lim_{x\to \infty}\frac{d \{\ln {x}\}/dx}{d\{x^{2}-1\}/dx}=\lim_{x\to \infty}\frac{1}{2x^{2}}=0.
\end{align}

To summarize, we have $0<G(x)<{1}/{2}$ for $x>1$. As such, ${\partial \{A(Q)\}}/{\partial Q}$ in \eqref{fod:A(Q)} can be expressed as
\begin{align}
\frac{\partial \{A(Q)\}}{\partial Q}
    & > \frac{\sqrt{T}\left[1-G(x)\right]}{\sqrt{(1+Q)^{2}-1}}
 > \frac{\frac{1}{2}\sqrt{T}}{\sqrt{(1+Q)^{2}-1}}>0.
\end{align}

Based on the sign of $\frac{\partial \{A(Q)\}}{\partial Q}$, we have $\frac{\partial \{\epsilon(Q)\}}{\partial Q}<0$ as per \eqref{fod:Epsilon}. Next, we derive $\frac{\partial^2  \{A(Q)\} }{\partial Q^2}$ in \eqref{sod:A(Q)}. As such, the analysis of the sign of $\frac{\partial^2  \{A(Q)\} }{\partial Q^2}$ is equivalent to determining the value of $\Psi_{1}+\Psi_{2}-\Psi_{3}-\Psi_{4}$. In order to address this problem, we decompose it into two parts, and calculate $\Psi_{1}-\Psi_{3}$ and $\Psi_{2}-\Psi_{4}$, respectively.
Firstly, we calculate $\Psi_{1}-\Psi_{3}$ as
  \begin{align}\label{compare:Psi1-Psi3General}
  \Psi_{1}-\Psi_{3}&<\bigl(3G(x)-2\bigr)/{\left(\sqrt{(1+Q)^{2}-1}\right)^{3}}.
  \end{align}
Similarly, $\Psi_{2}-\Psi_{4}$ can be obtained as
  \begin{align}\label{compare:Psi2-Psi4General}
  \Psi_{2}-\Psi_{4}&< \bigl(3G(x)-1\bigr)/{\sqrt{(1+Q)^{2}-1}}.
  \end{align}
If we have $\eqref{compare:Psi1-Psi3General}<0$ and $\eqref{compare:Psi2-Psi4General}<0$ simultaneously, we can guarantee $\frac{\partial^2  \{A(Q)\} }{\partial Q^2}<0$. We note that $\eqref{compare:Psi1-Psi3General}<0$ and $\eqref{compare:Psi2-Psi4General}<0$ are equivalent to $3G(x)-2<0$ and $3G(x)-1<0$. Thus, we only need to ensure $3G(x)-1<0$ due to $0<G(x)<\frac{1}{2}$ for $x>1$.
Now, we substitute $G(x)=\frac{\ln {x}}{x^{2}-1}=\frac{\ln(1+Q)}{(1+Q)^{2}-1}$ into $3G(x)-1<0$, and we have
  \begin{align}\label{compare:final_step}
  &3G(x)-1 <0
  \Longrightarrow \frac{\ln(1+Q)}{(1+Q)^{2}-1} < \frac{1}{3}
  \overset{(a)}{\Longrightarrow} Q>\textrm{max}(Q_0,0).\notag
  \end{align}
where $\overset{(a)}{\Longrightarrow}$ is obtained due to the fact that $G(Q)$ is a decreasing function w.r.t. $Q$ for $Q>0$, where $Q_0$ is the solution to ${\ln(1+Q)}/{(1+Q)^{2}-1}={1}/{3}$.

So far, we prove that $\Psi_{1}+\Psi_{2}-\Psi_{3}-\Psi_{4}<0$ for $Q >Q_0$. As a result, we have $\frac{\partial^2  \{A(Q)\} }{\partial Q^2}<0$ when $Q>Q_0$.
To summarize, for $Q>Q_0$, we have $\frac{\partial^2 \epsilon(Q)}{\partial Q^2}>0$ in \eqref{sod:Epsilon} due to $A(Q)>0$ and $\frac{\partial^2 \{A(Q)\} }{\partial Q^2}<0$.
Thus, we have $\frac{\partial^2 P_{\epsilon}(Q) }{\partial Q^2}>0$ for $Q_0  < Q < P_{\mathrm{max}}(N_{\mathrm{A}}-1)$, which completes the proof.



\end{document}